# $B$ Physics with NRQCD: A Quenched Study

Presented by A. Ali Khan [a*]

[a]Department of Physics and Astronomy, University of Glasgow, UKQCD Collaboration

We present results on the spectrum of $B$ mesons and heavy baryons, using a non-relativistic formulation for the heavy and a clover action for the light quark. We also discuss $B$ meson decay constants and their dependency upon the heavy meson mass.

## 1. INTRODUCTION

For some years, particle theorists have put considerable effort into calculations of the $B$ meson decay constant $f_B$. This quantity is required, together with the bag parameter $B_B$, for a determination of the CP-violating phase in the CKM matrix. However, relativistic quark methods need some reinterpretation if they are to work for $b$ quarks [1]. Simulations are generally done at lighter quark masses and results extrapolated to the $b$, assuming an expansion in the inverse heavy quark mass, as predicted by Heavy Quark Effective Theory (HQET), up to quadratic terms. Some difficulties in matching with the static results are often encountered. Extrapolation has also to be performed to obtain the spectrum around the $B$, and it is hard to get sensible values for the hyperfine splitting (a successful static calculation is in [2]). However, with NRQCD it is possible to simulate in mass regions around and above the $b$ quark. NRQCD is an effective theory which systematically includes higher dimensional operators in an expansion in the inverse heavy quark mass $1/m_Q^0$. One can study contributions of operators occuring in HQET separately. We are performing a calculation of the spectrum and decay constants of heavy-light mesons. Our study of the spectrum comprises $S$ and $P$ wave states, as well as the $\Lambda_b$ baryon. $P$ states have recently been discovered experimentally and provide a way to determine the lattice spacing from the $B$ system itself.

*In Collaboration with S. Collins (SCRI), C. T. H. Davies, (Glasgow Univ., UKQCD collab.), J. Shigemitsu (Ohio State Univ.) and J. Sloan (SCRI).

## 2. THE COMPUTATION

The NRQCD Lagrangian in this calculation and in [3] is corrected through $O(1/m_Q^0)$:

$$\mathcal{L} = Q^\dagger \left( D_t - \frac{\vec{D}^2}{2m_Q^0} \right) Q - Q^\dagger \frac{\vec{\sigma}\vec{B}}{2m_Q^0} Q, \qquad (1)$$

where $Q$ is the Pauli spinor describing the heavy quark. The gauge links are tadpole improved:

$$U_\mu \rightarrow U_\mu / u_0, \;\; u_0^4 = \langle 1/3 Tr U_{\text{Plaq.}} \rangle, \qquad (2)$$

so we use the tree level coefficients for the $1/m_Q^0$ terms in (1). For the calculation of the heavy-light decay constants it is important to also include the $1/m_Q^0$ corrections to the currents. These can be obtained by doing an inverse Foldy-Wouthuysen transformation on the heavy quark spinor:

$$q_h = \left( 1 - \frac{\vec{\gamma}\vec{D}}{2m_Q^0} \right) Q. \qquad (3)$$

Thus we calculate the decay constants from matrix elements $\langle 0|C|M\rangle$, where

$$C = \bar{q}_h \Gamma Q - \frac{1}{2m_Q^0} \bar{q}_h \Gamma (\vec{\gamma}\vec{D}) Q \qquad (4)$$

is given by the time component of the axial vector current ($M = B$, $\Gamma = \gamma_5\gamma_0$) or the vector current ($M = B^*$, $\Gamma = \gamma_\mu$), respectively.

We simulate heavy quarks with masses $m_Q^0 =$ 1.71, 2.0, 4.0, 8.0 and in the static approximation. We use light quarks with a tadpole improved clover action at $\kappa$ values 0.1370 and 0.1381. The calculation is being done on quenched configurations, generated by UKQCD, at $\beta = 6.0$, fixed to Coulomb gauge, on a lattice volume of $16^3 \times 48$. The results presented here are preliminary, based on an ensemble of ca. 50 configurations at $m_Q^0 = 8.0$ and ca. 60 configurations at the other heavy quark masses. Using 130 configurations the lattice spacing has in the calculation of [4] been determined from light spectrospcopy ($m_\rho$) to be $a^{-1} = 1.891(103)$ GeV and $\kappa_c$ to be 0.139245(33). These data are compared to a previous calculation with clover light fermions with clover coefficient $c = 1$ (i.e. not tadpole improved) at $\kappa = 0.1432$ and 0.1440 and heavy masses 1.71, 2.0, 2.5, 4.0 at the same $\beta$ value and lattice volume, with a lattice spacing from light spectroscopy of 2.05(10) GeV. In the quenched approximation experience shows that $a$ depends upon the system used to determine it, so it is desirable to use a value from the $B$ system itself, like the $S-P$ splitting. Until sufficiently accurate simulation results for this are available, we use the values from light spectroscopy for our heavy-light results. We expect the light quark momentum to set the energy scale in $B$ mesons. The $\kappa$ values are tuned such that the heavier and the lighter ones of each give approximately the same $\pi$ mass. Implemented are $^1S_0$, $^3S_1$, $^3P_0$, $^3P_1$, $^3P_2$ and $^1P_1$ states, taking also cross correlations between $P$ states into account. Smearing functions are hydrogen-like ground and 1st excited state wave functions. For each meson we calculate all combinations of smeared and local operators at source and sink. The $\Lambda_b$ baryon is smeared at the source and local at the sink.

Table 1
Bare ground state energies (heavier $\kappa$).

| $am_Q^0$ | $a\Delta E_0$ | |
|---|---|---|
| | $c=1$ | $c=1/u_0^3$ |
| 1.71 | 0.501(9) | 0.516(5) |
| 2.0 | 0.506(6) | 0.520(4) |
| 2.5 | 0.510(6) | |
| 4.0 | 0.511(7) | 0.532(7) |
| 8.0 | | 0.537(7) |
| static | 0.524(6) | 0.542(8) |

## 3. ANALYSIS AND RESULTS

### 3.1. B mass and decay constants

To extract ground state energies ($E_0$) and amplitudes we fit smeared-local ($C_{SL}$) and smeared-smeared ($C_{SS}$) correlation functions simultaneously to a single exponential. For sufficiently large times one has:

$$C_{SS} \to Z_S^2 e^{-aE_0 t}, \quad (5)$$

$$C_{SL} \to Z_S Z_L e^{-aE_0 t}. \quad (6)$$

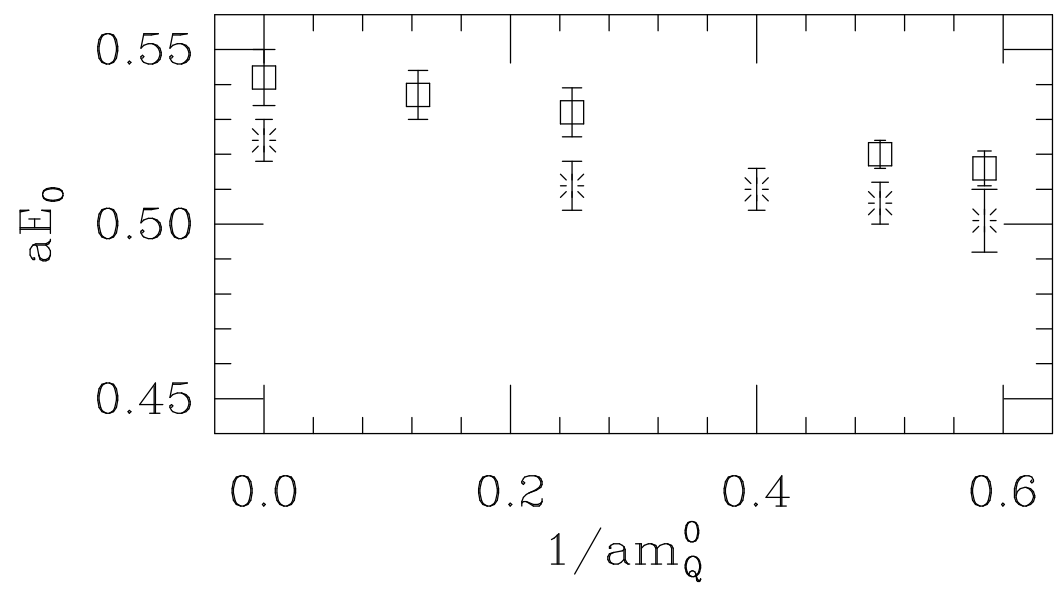

Figure 1. Ground state energies for tadpole improved clover fermions (squares, $\kappa = 0.137$) and clover fermions with c = 1 (bursts, $\kappa = 0.1432$).

Having extracted $Z_L$, we can write down the matrix elements. For the pseudoscalar meson we have $\sqrt{2}Z_L = a^{3/2} f_P \sqrt{M_P}$ and for the vector meson $\sqrt{2}Z_L = a^{3/2} f_V \sqrt{M_V}$. The meson mass we obtain from the relation

$$M = E_0 + \Delta. \quad (7)$$

$\Delta$ includes the renormalized heavy quark mass and the divergent zero point energy of the quark. The values used in this work are calculated by C. Morningstar [5]. Ground state energies for the heavier $\kappa$ values are listed in table 1. We find that the results are larger for the tadpole improved than for the non-tadpole improved light fermions, as shown in figure 1. This is expected from a perturbative analysis of the effect of increasing the coefficient of the clover term.

Table 2
Meson masses and decay constants (tadpole improved light fermions, $\kappa = 0.1370$), all values in lattice units.

| $m_Q^0$ | $M_P$ | $f_P\sqrt{M_P}$ | $f_V\sqrt{M_V}$ | spin avg. |
|---|---|---|---|---|
| 1.71 | 2.25 | 0.177(5) | 0.200(5) | 0.194(4) |
| 2.0 | 2.54 | 0.186(5) | 0.208(7) | 0.188(5) |
| 4.0 | 4.60 | 0.220(7) | 0.232(8) | 0.229(78) |
| 8.0 | 8.17 | 0.253(8) | 0.260(10) | 0.258(9) |
| static | $\infty$ | 0.296(9) | | |

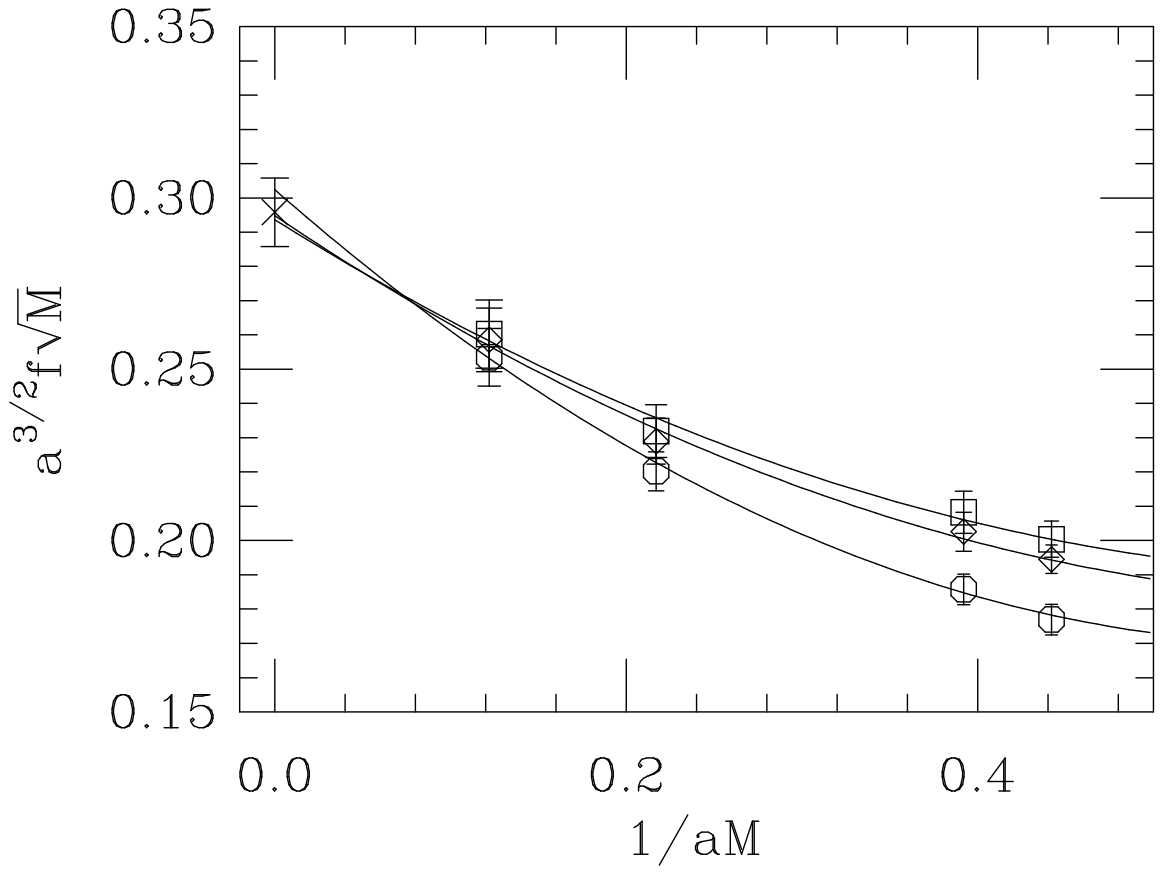

Figure 2. Lattice matrix element $a^{3/2}f\sqrt{M}$ with NRQCD and tadpole-improved light fermions (circles: pseudoscalar, squares: vector, diamonds: spin averaged, cross: static) plotted against the inverse spin averaged meson mass. The lines are correlated fits to the NRQCD points.

Table 3
Meson masses and $f_P$ in GeV.

| | $c=1$ | | $c=1/u_0^3$ | |
|---|---|---|---|---|
| $am_Q^0$ | $M_P$ | $f_P$ | $M_P$ | $f_P$ |
| 1.71 | 4.5(5) | 0.16(4) | 4.1(5) | 0.16(3) |
| 2.0 | 5.1(6) | 0.16(4) | 4.7(6) | 0.16(3) |
| 2.5 | 6.0(8) | 0.18(3) | | |
| 4.0 | 9.3(7) | 0.19(5) | 8.6(11) | 0.18(4) |
| 8.0 | | | 15.4(13) | 0.21(3) |
| static | | 0.28(5) | | 0.22(6) |

Thinking of the light quarks non-relativistically, the clover term adjusts the effective Darwin term in the action and this shifts energies upwards [6]. The $B/\rho$ ratio is lower for the tadpole improved fermions.

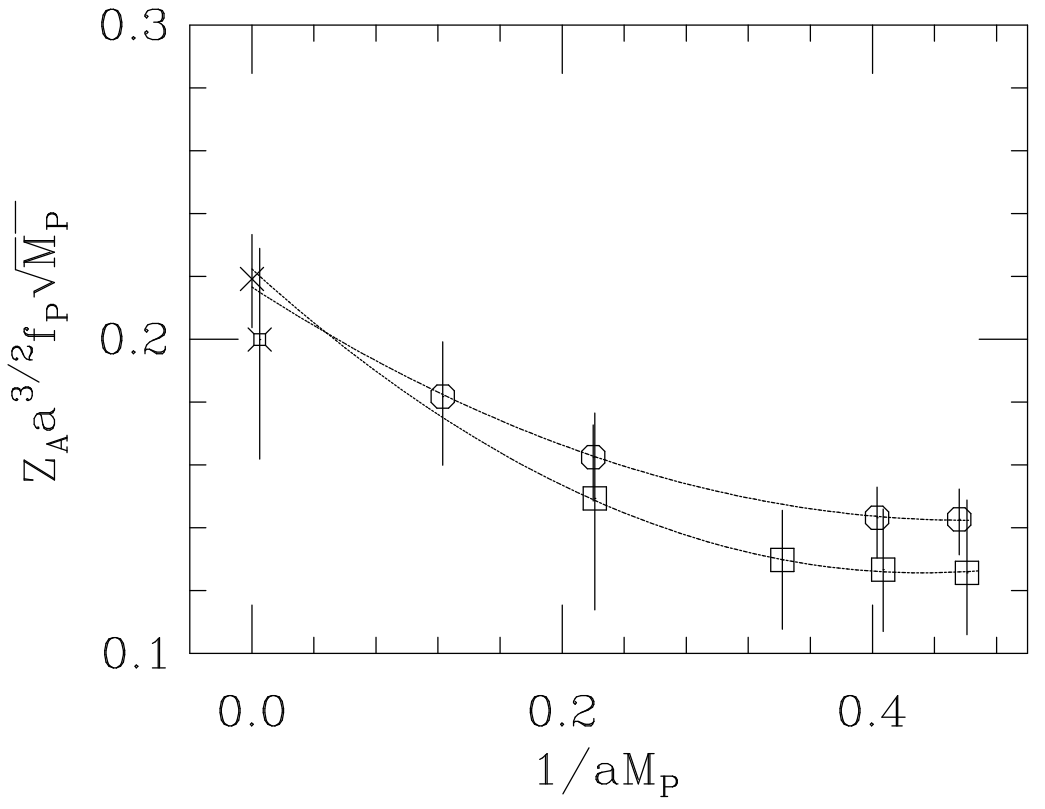

Figure 3. Chirally extrapolated pseudoscalar matrix element $Z_A a^{3/2} f_P\sqrt{M_P}$ with tadpole improved light fermions (circles: NRQCD, fancy cross: static) and light fermions with c = 1 (squares: NRQCD, cross: static) and correlated fits to the NRQCD points.

The lattice results for the pseudoscalar, vector and spin averaged decay constants are given in table 2. The extrapolation of the NRQCD points is compatible with the static simulation result, as seen in figure 2. This remains true if one includes the matching renormalization constants from the lattice to the continuum effective theory. The slope increases for large masses. It is considerably higher than previous lattice calculations and of the order of QCD sum rule predictions [7]. This, along with an estimate of the contributions of the various operators at $O(1/m_Q^0)$, is discussed in more detail in [3]. To convert the decay constants into physical units, we are calculating perturbatively the matching renormalization constants to the full theory. A comparison of the chirally extrapolated pseudoscalar matrix elements $Z_A f_P\sqrt{M_P}$ for $c=1$ and $c=1/u_0^3$ clover light fermions is shown in figure 3. Chirally extrapolated values for $f_P$, together with the corresponding meson masses, are given in table 3. For the $B$ meson we take 0.16(4) GeV as our estimate for the quenched decay constant. The major contribution to this error comes from the uncertainty in $a$.

### 3.2. Mass splittings

The mass splitting $\Delta E$ between hadrons is extracted by fitting the ratio of the corresponding correlation functions to a single exponential:

$$\frac{C_{SL}^{(1)}}{C_{SL}^{(0)}} \propto e^{-a\Delta Et}, \tag{8}$$

where $C_{SL}^{(0)}$ denotes the ground state smeared-local correlation function of the $^1S_0$ state and $C_{SL}^{(1)}$ the $^3S_1$ correlation function (for the hyperfine splitting) or the $\Lambda_b$ correlation function (for the $\Lambda_b$−B splitting).

#### 3.2.1. $\mathbf{B^* - B}$ splitting

The fit results for the $B^* - B$ splitting at $\kappa = 0.1370$ in lattice and in physical units are given in table 4. For the lighter $\kappa$ the statistical errors are considerably larger; there is no visible $\kappa$ dependence. So we approximate the splitting by our results at $\kappa = 0.1370$. From HQET we expect the splitting to have the following dependence on the heavy meson mass:

$$\Delta E \propto 1/M_P. \tag{9}$$

Performing a correlated fit of the simulation results to a linear function, we find that their behaviour is compatible with this expectation. The splittings in the mass region of the B are slightly lower than the experimental value of 0.46(1) MeV, which might be an effect of quenching.

#### 3.2.2. $\mathbf{\Lambda_b - B}$ splitting

The chirally extrapolated results for the $\Lambda_b - B$ are listed in table 4. Our splitting is compatible with the experimental value 0.36(5) GeV. The central value is slightly high, but this might be due to contamination with excited states, since correlation functions get large errors for larger times.

## 4. CONCLUSIONS

This is a report on an ongoing computation of the $B$ meson spectrum and decay constants with NRQCD in the quenched approximation. We compare results using tadpole improved clover light fermions with $c = 1$ clover light fermions. The tadpole improved results turn out to be subject to stronger statistical fluctuations. One observes that the tadpole improved ground state energies are slightly enhanced. Now also a very high quark mass is included into the simulation, and we find that the extrapolation of the NRQCD decay constants to infinite mass is compatible with the static value. The matching renormalization constants to the full theory are being calculated in NRQCD perturbation theory. Using preliminary perturbative results we quote a value of $f_B = 0.16(4)$ GeV for the $B$ decay constant. NRQCD also provides us with a tool for computing $B^* - B$ and $\Lambda_b - B$ splittings at the $B$. The hyperfine splitting is proportional to $1/M_P$ as predicted by HQET. In the region of the $B$ mass, we find a splitting slightly lower than experiment. The $\Lambda_b - B$ splitting is compatible with the experimental value.

Table 4
Hyperfine and baryon-meson splittings (clover term $c = 1/u_0^3$).

| | $\Delta E(B^* - B)$ | | $\Delta E(\Lambda_b - B)$ | |
|---|---|---|---|---|
| $m_Q^0$ | $a\Delta E$ | $\Delta E$[MeV] | $a\Delta E$ | $\Delta E$[GeV] |
| 1.71 | 0.020(1) | 38(6) | 0.25(7) | 0.47(15) |
| 2.0 | 0.017(1) | 32(6) | 0.25(7) | 0.48(15) |
| 4.0 | 0.0091(8) | 17(3) | 0.27(7) | 0.50(16) |
| 8.0 | 0.0042(7) | 8(2) | 0.26(7) | 0.50(16) |

This work was supported by SHEFC, PPARC, the U.S. DOE and the NATO under grant number CRG 941259.